\documentclass{JHEP3}
\usepackage{epsfig}
\usepackage{graphicx}% 
 
\def\lesssim{\mathrel{\hbox{\rlap{\hbox{\lower4pt\hbox{$\sim$}}}\hbox{$<$}}}}
\def\gtrsim{\mathrel{\hbox{\rlap{\hbox{\lower4pt\hbox{$\sim$}}}\hbox{$>$}}}}

\title{Caustics in Tachyon  Matter and Other\\Born-Infeld Scalars}

\author{Gary Felder\\
CITA, University of
Toronto, 60 St. George Street, Toronto, ON M5S 3H8, Canada\\ E-mail: \email{felder@cita.utoronto.ca}} \author{Lev Kofman\\
    CITA, University of
Toronto, 60 St. George Street, Toronto, ON M5S 3H8, Canada\\ E-mail: \email{kofman@cita.utoronto.ca}} \author{Alexei Starobinsky\\
    Landau Institute for Theoretical Physics, Kosygina 2, Moscow, 119334, Russia\\
    E-mail: \email{alstar@landau.ac.ru}
}
\received{\today}        
\preprint{CITA-02-22}
 
\abstract{We consider scalar Born-Infeld type theories with arbitrary potentials $ V(T)$ of a scalar field $T$. We find that for models with runaway potentials $V(T)$ the generic inhomogeneous solutions after a short transient stage can be very well approximated by the solutions of a Hamilton-Jacobi equation that describes free streaming wave front propagation. The analytic solution for this wave propagation shows the formation of caustics with multi-valued regions beyond them. We verified that these caustics appear in numerical solutions of the original scalar BI non-linear equations. Our results include the scalar BI model with an exponential potential, which was recently proposed as an effective action for the string theory tachyon in the approximation where high-order spacetime derivatives of $T$ are truncated. Since the actual string tachyon dynamics contain  derivatives 
of all orders, the tachyon BI model with an exponential potential becomes inadequate when the caustics develop because high order spatial derivatives of $T$ become divergent.\\
BI type tachyon  theory with   a potential   decreasing at large $T$
could have interesting cosmological applications because the tachyon field 
rolling towards its ground state at infinity acts as pressureless dark matter. We find that inhomogeneous cosmological tachyon fluctuations  rapidly grow 
and develop multiple caustics.  Any considerations of the role of the tachyon field in cosmology will have to involve finding a way to predict the behavior of the field at and beyond these caustics.}

\keywords{tac.sft.dbr}

\begin{document}

\section{Introduction}

In this paper we investigate generic  solutions of scalar Born-Infeld field theories of the form \begin{equation}\label{action} S = -\int d^Dx \, V(T) \sqrt{1 + \alpha' \partial_\mu T \partial^\mu
  T} \ ,
\end{equation}
where $T(x^{\mu})$ is a (dimensionless) scalar field and $V(T)$ is its potential. The dimensional parameter  $\alpha'$ has units of $mass^{-2}$. In string theory $M_s=l_S^{-1}=1/\sqrt{\alpha'}$ are the fundamental string mass and length scales. For the remainder of the paper we will work in units where $\alpha'=1$.

Although this problem has merit by itself, especially as a BI generalization of the sigma model approach to the effective string theory action \cite{sigma}, the reason to explore this theory now is related to  the rolling tachyon and its application to  cosmology.  Recently Sen conjectured \cite{Sen1,Sen2} that the qualitative dynamics of a string theory tachyon field in a background of an unstable D-brane system could be approximated with the effective action (\ref{action}) with
$V(T) \sim e^{-T}$.

However, string theory tachyon calculations are  reliable only in the approximation where derivatives $\nabla^{\mu} T$ are truncated beyond the quadratic order \cite{trun}. For instance, the 4D effective field theory action of the tachyon field $T$ on a D$3$-brane, computed in bosonic theory around the top of the potential, up to higher derivative terms, is given by \cite{trun} \begin{equation}\label{action1} S_B = \tau_3\int d^4x\, \sqrt{g}\left(\alpha' e^{-T}  \partial_\mu{T}\partial^\mu{T} + (1+T) e^{-T}\right) + O(  \partial_\mu \partial^\mu{T}) \ , \end{equation} where $\tau_3$ is the brane tension.  It is far from clear that the model (\ref{action}) describes actual tachyon dynamics. Still, this relatively simple formulation of tachyon dynamics (\ref{action}) triggered significant interest in the investigation of the role of tachyons in cosmology \cite{cosm,FKS,SW,KL,STW} and in the field theory of tachyons \cite{Sen3}. . Even before the recent works on tachyon cosmology, a model known as ``k-inflation'' \cite{k}, based on the action (\ref{action}), had been considered in cosmology in the search for interesting equation of states. 
Note that  the BI model (\ref{action}) with a constant potential $V=1$ and a time-like derivative of the tachyon field ($\partial_{\mu}T \partial^{\mu}T < 0$) is equivalent to the Chaplygin gas model that was proposed as a model for both dark matter and dark energy in the present Universe~ \cite{Kam}.

In what follows we will
continue to call the scalar field $T$ a tachyon field, bearing in mind that the connection of this field to the actual string theory tachyon is uncertain and that it might also have application to a broad potential pool of other theories.

Our investigation combines both analytic approximations and ``heavy duty'' field theory lattice simulations \cite{latticeasy}, which we adopt here for the theory (\ref{action}). We found that for BI scalars with runaway potentials $V(T)$, i.e. potentials for which $T \rightarrow \infty$ over time, the most interesting features of the inhomogeneous solutions of the equations of motion are the formation of caustics and the folding of the hypersurface $T(t,\vec{x})$.

In section \ref{eqofmotion2} we consider the equation of motion for the tachyon field in Minkowskii spacetime in $1+1$ dimensions. In this simple situation we are able to clearly show our main results. (In later sections we will show that these same results apply to $3+1$ dimensions and  an expanding universe.) First we consider an exponential potential $V(T) \sim e^{-T}$, and we assume some initial inhomogeneous realization of the tachyon field $T(x)$.  In cosmology this is usually a realization of a random Gaussian field. For 
theories where $\alpha'$ is a fundamental scale,
  consistency of the effective field theory
 action approach requires
that the gradients (in fundamental units)
not be large, $\nabla_x T \lesssim 1$.  
 Assuming initial  spatial gradients are not large,
 we argue analytically and
confirm numerically that after a quick, transient regime, the equation of motion reduces to a simple first order Hamilton-Jacobi equation  that describes  the propagation of free massive  wave fronts. We then show that this same approximation holds for a wide class of runaway potentials.

In section \ref{rootarg} we solve this Hamilton-Jacobi equation  exactly using the method of characteristics and find that the solution quickly develops caustics. We suggest a couple of geometrical techniques to view the time evolution of $T$ during this process. We also solve the original equation of motion numerically for a realization of a Gaussian field and show the formation of caustics from the regions of small spatial gradients, in agreement with the results of our approximation. 

In section \ref{dim} we extend the results on caustics in  the $T(t, \vec x)$ field to $3+1$ dimensions. The qualitative results are unchanged.

In section \ref{eqofmotion1} we discuss the cosmological implications of our results. The cosmological applications of a tachyon field $T$ with an exponential potential are two-fold. First, a time dependent field $T$ with a sufficiently rapidly decreasing potential ($T^2V(T)\to 0$ at $T \to \infty$) has a pressureless ``cold 
dark matter'' equation of state in the late time asymptotics 
$t \gg 1$ with its energy density decreasing as $\varepsilon \propto a^{-3}$, 
where $a(t)$ is the scale factor of a Friedmann-Robertson-Walker
(FRW) cosmological model of the Universe \cite{Sen2,FKS}. If the energy density of the rolling tachyon becomes dominant, it could be a dark matter candidate. Second, the tachyon appears in various models of brane inflation. In models with unstable $ D$-branes or annihilating $ D$ and $\bar D$ branes where the unstable tachyon mode has a ground state at $ T \to \infty$, ironically, it is exactly the above property of the tachyon that poses a problem. Indeed, the energy density of the pressureless tachyon almost always ends up dominating \cite{KL,STW}.

A consistent investigation of the tachyon field in cosmology must include spatial inhomogeneities. The tachyon potential has a maximum with negative curvature. As it rolls from this maximum inhomogeneities will be generated from its vacuum fluctuations, as in the theory of spontaneous symmetry breaking (tachyonic preheating) \cite{preheating}.

The evolution of linear fluctuations of a rolling tachyon in an expanding universe was considered in \cite{FKS,SW}. As would be expected, tachyon fluctuations coupled with small metric fluctuations grow with time due to gravitational instability, just like usual pressureless matter. However, the linear stage of tachyon fluctuations very quickly changes to a strongly non-linear stage.  In the usual cold dark matter scenario cosmologists understand well the non-linear bottom-up evolution of gravitational instability.  In particular, the theory of non-linear structure formation includes caustics and multi-streaming of collisionless dark matter particles. (For CDM, however, early small scale caustic formation is a  transient phenomenon).  Many qualitative aspects of this theory can be understood even without expansion of the universe in the much simpler model of media of cold collisionless particles with a smooth initial potential velocity field \cite{ASZ}.  We shall address a similar problem for  the tachyon model (\ref{action}).

Thus, to pursue tachyon cosmology further, we have to understand the non-linear evolution of inhomogeneities, i.e. the generic solution of (\ref{action}). In section \ref{eqofmotion1} we write down the equation of motion for the tachyon field in an expanding universe and recall the results of \cite{FKS} regarding linear 
growth of tachyon fluctuations. 
 We develop an analysis of non-linear inhomogeneities of the tachyon in an 
expanding universe. Again, the equation for the tachyon field is reduced to the covariant form of the Hamilton-Jacobi equation.  We argue that the fully non-linear evolution of tachyon dark matter leads to the formation of caustics and multi-valued regions beyond them, similar to what we derived in the simpler case of sections \ref{eqofmotion2}-\ref{dim} without expansion of the universe.

In section \ref{discussion} we discuss the issue of how to physically interpret the development of caustics. Formally, caustics border the regions where the solution becomes multi-valued. The physical interpretation of multi-valued solutions and caustics depends on the scalar field theory  specified by  the potential $V(T)$ and the dimensions of space-time. Second order and higher derivatives blow up on caustics. Therefore the action (\ref{action}), which is supposed to work in the approximation of truncation of
high order derivatives    ,
 is no longer a valid approximation to string theory tachyon dynamics once caustics have formed. It is unclear how to work with multi-valued solutions $T$ in the four dimensional theory derived for a single-valued effective field $T$.

Also, the interaction of the scalar BI field with other fields (like  the dilaton or gauge fields in the tachyon case) may be important at caustics. We discuss some of the issues involved in making such an interpretation, but we do not 
intend to resolve
the question here.

\section{Inhomogeneous Non-Linear Tachyon Field in $1+1$ Flat Spacetime}\label{eqofmotion2}

The field equation derived from (\ref{action}) is \begin{equation}\label{motion1} \nabla_\mu\nabla^\mu T - \frac{\nabla_\mu \nabla_\nu T} {1 + \nabla_\alpha T \nabla^\alpha T} \,\, \nabla^\mu T \,\, \nabla^\nu T
 - \frac{V_{,T}}{V}=0 \ ,
\end{equation}
where $\nabla_{\mu}$ is a covariant derivative with respect to the metric $g_{\mu\nu}$. The energy momentum tensor of tachyon matter is \begin{equation}\label{tmunu} T_{\mu\nu}=\frac{V}{\sqrt{1+\nabla_{\alpha}T \nabla^{\alpha}T}} \, \nabla_{\mu}T \, 
\nabla_{\nu}T -g_{\mu\nu} V \, \sqrt{1+\nabla_{\alpha}T \nabla^{\alpha}T} \ . \end{equation} For the rest of this section we will consider a tachyon field in $1+1$ dimensional Minkowskii space. We will discuss generalizations of our results to higher dimensions and expanding universes in later sections.

\subsection{The Equation of Motion}

Specializing to $1+1$ dimensional flat space,  equation (\ref{motion1}) becomes \begin{equation}\label{1dequation}
\ddot{T} = \left(1 + T'^2\right)^{-1} \left[\left(1
-\dot{T}^2 + T'^2\right) \left(T'' - \frac{V_{,T}}{V} \right) + 2 T' \dot{T} \dot{T}' - T'^2 T''\right] \ , \end{equation} where dots and primes indicate time and space derivatives respectively.

First we consider  an exponential potential
\begin{equation}
V(T) = V_0 e^{-T/T_0} \ ,
\end{equation}
but later we also will discuss other potentials $V(T)$.  We can without loss of generality set $T_0=1$ by appropriate rescalings of  spacetime coordinates, which will be measured in units of $T_0$.

\subsection{The Free Streaming Approximation for the Exponential Potential}\label{approx}

For a homogeneous field with $T'=0$ equation (\ref{1dequation}) can be solved exactly to give \begin{eqnarray} T &=& \ln\left(\cosh(t-t_*)\right) + T_* \\ \dot{T} &=& \tanh(t-t_*), \end{eqnarray} where $t_*$ and $T_*$ give the time at which $\dot{T}=0$ and the value of $T$ at that time respectively. Asymptotically $\dot{T}$ approaches $1-2e^{-2t}+O(e^{-4t})$  and $T(t) \simeq t+e^{-2t}+ O(e^{-4t})$. 

For an inhomogeneous field the situation seems at first glance to be much more complicated.  Indeed, equation (\ref{1dequation}) is a non-linear partial differential equation. However, we can give some general arguments for how the field should behave and then check our conclusions numerically.  Let us assume that the gradients are not large, $T', T'' < 1$. Solving equation (\ref{1dequation}) numerically, we found that the field rapidly approached the state $1 - {\dot T}^2 + T'^2=0$. This means that $T(t, x)$ is very close (exponentially in fact) to some auxiliary field that we denote as $ S(t, x)$ and that satisfies the equation \begin{equation}\label{free} \dot{S}^2 - S'^2 = 1 \ . \end{equation} Because this combination will appear many times in the analysis that follows we will define the operator  $P(T) \equiv 1 - {\dot T}^2 + T'^2$. Thus, equation (\ref{1dequation}) has an attractor $P(T)=0$. (Note that if we multiply equation (\ref{1dequation}) by $P(T)$, then $S$ will be a {\em partial} solution of the resulting equation.)

One way to get a feeling for why the complicated equation
(\ref{1dequation}) reduces to the very simple form (\ref{free}) is the following. Let us rewrite (\ref{1dequation}) without the last two terms in its {\it r.h.s.}, and dropping $T'^2$ in the factor $(1+T'^2)$.  We have \begin {equation}\label{drop} \ddot T = \left( 1-\dot T^2 + T'^2 \right) (T''+1) \ . \end{equation} Assuming $T'' > -1$, this equation rapidly approaches the state $P(T)=0$.  Indeed, if initially $P(T)=\left( 1-\dot T^2 + T'^2 \right)> 0$, then $\ddot T >0$ and $\dot T$ increases, which results in $P(T)$ asymptotically approaching zero from above.  If initially $P(T)<0$ , then $\ddot T <0$, and $P(T)$ asymptotically approaches zero from below.

We did not pursue an investigation of precisely how small the initial values of $T'$ and $T''$ need to be for this approximation to be valid; we just assume that in dimensional units $T' < l_s^{-2}$, $T'' < l_s^{-3}$.  However, by using effective field theory we are implicitly assuming that scales of inhomogeneities are greater than $l_s$, so these inequalities must hold.  Also, for realistic scenarios of tachyon fields in cosmology they are definitely small. Indeed, initially $T' \sim {l_s \over L} {T_i \over T_0} \ll 1$, where $L$ is the scale of cosmological ingomogeneities, which is always expected to be larger than $l_s$.

For applications of the action (\ref{action}) other than string theory, however, one might imagine that $\alpha' \gg M_s^2$ and thus the initial gradients could in principle be large. We note, however, that for any initial configuration $T(x)$ for which there are minima there will always be some neighbourhood around each minimum for which our approximation is valid. Of course the gradients are also small around the maxima of the field configuration, but as we will see later it is in regions around the minima where most interesting effects -- caustics -- develop.

We have numerically tested a variety of initial conditions and found that for $T' \lesssim 1$  the field rapidly asymptotes to $P = 0$ everywhere. For configurations with large gradients we have found that $P$ rapidly approaches $0$ in the neighbourhood of the extrema of the initial configuration.

\subsection{The Free Streaming Approximation for Other Potentials $V(T)$}\label{other}

The form (\ref{free}) describes the large value asymptotic solution of equation (\ref{1dequation}) without any trace of the potential $V(T)=V_0 e^{-T/T_0}$. Therefore it is natural to ask whether this asymptotic solution may work for other potentials. To this end we consider runaway potentials $V(T)$ where asymptotically $T \to \infty$.
 Such runaway potentials include
$V(T)=e^{-T^2}$ and $V(T)=1/T^2$, which appear in the literature on cosmological tachyon matter \cite{cosm}. \footnote{However, below we will 
see that $V(T)=T^{-2}$ is the 
 borderline case for the free streaming approximation in an {\em expanding} Universe in $3+1$ dimensions.}  We checked numerically that solutions of equation (\ref{1dequation}) with these potentials have large value asymptotics of $T$ that are described by the equation (\ref{free}). We conjecture that {\it the solutions of all theories
(\ref{action}) with non-constant potentials $V(T)$ that admit runaway evolution of the scalar field $T$ can be approximated by the solutions of the free wave propagation equation (\ref{free}).} Another words, the solution of (\ref{free}) is an attractor of the equation (\ref{1dequation}). 

If the potential $V(T)$ admits only finite valued evolution of the scalar field $T$, the solution (\ref{free}) may or may not appear as an intermediate asymptotic, depending on the form of the potential $V(T)$. Certanly, if the potential $V(T)$ has a minimum at $T \sim T_0$, the Hamilton-Jacobi asymptotic solution does not work. 

It is interesting to consider the marginal case $V(T)=1$. This case corresponds to the classical scalar BI theory. Consider a toy model where $T$  describes transversal oscillations of a light $3+1$ dimensional brane embedded in five dimensional space-time.  This case is essentially different from the action (\ref{action}) with an exponential potential because the term $V_{,T}$ vanishes in the equation of motion. Indeed, the equation of motion of the scalar BI actions with constant $V$ admit plane wave solutions and their
(multi-valued) superpositions.  By proper coordinate transformation, the equation of motion can be reduced to a free scalar field wave equation that can be solved analytically, see e.g. \cite{BC}. In general, this wave equation has solutions with caustics and multi-valued regions beyond caustics (similar to that, say, in classical wave optics). We used some of the analytic solutions in \cite{BC} to check our numerical code for the general equation (\ref{1dequation}). \footnote{Any reader who is interested in the numerical solution for this particular case $V=1$ can find movies illustrating two and more waves scattering at {\tt <http://cita.utoronto.ca/${\sim}$felder/borninfeld>}. Conservations laws  evident in the scattering reflect the large symmetry of  the theory with $V=1$.}

\section{Exact Solutions in the Free Streaming Approximation}\label{rootarg}

The advantage of reducing equation (\ref{1dequation}) to the simple form (\ref{free}) is that the latter can be treated analytically. \footnote{An independent derivation of this analytic solution was discovered by \cite{GKS}.} We return to the notation $T$ for the scalar field, assuming it is decribed now by the equation (\ref{free}).  The 
analytic solution corresponds to free relativistic massive  wave propagation along characteristics. It also admits a simple geometric interpretation of how to construct ``wave front'' hypersurfaces $T(t, x)$ at any time $t$ directly from the initial conditions, without step-by-step integration of the wave equation. The most important feature of the free wave propagation, however, is the formation of caustics, where higher order gradients of the scalar field such as $T''$ blow up.

\subsection{The Solution in Characteristics}

In this section we will derive the analytic solution of equation
(\ref{free}) for the field $T(t, x)$ using the method of characteristics.  This equation has the form of a Hamilton-Jacobi equation for the action field $S$, which can be viewed as a hypersurface in the three dimensional auxiliary space $(T, t, x)$. The initial value is given by the hypersurface $T(t_0, q)$, where $t_0$ is the initial time and $q$ is the initial spatial coordinate, $x(t_0)=q$. The parameter $q$ is often called the Lagrangian coordinate, while $x$ is called the Eulerian coordinate. We will set $t_0=0$.

Equation (\ref{free}) can be viewed as a description of the free streaming of massive particles of mass $M_s$. The characteristics $x(t, q)$ in this case are trajectories of individual particles marked by their initial positions $q$. The vectors that are orthogonal to the family of hypersurfaces $S(t, x)$ are tangential to the family of characteristics $x=x(t, q)$.

Expressing this particle viewpoint mathematically, the method of characteristics gives us solutions along a family of curves parametrized by two variables, $q$ -- which defines which curve we are on, and $s$ -- an affine parameter along each curve. For initial conditions we set $x(q,0)=q$, $t(q,0)=0$, and $T(q,0)=T_i(q)$. We then have the characteristic equations \begin{eqnarray} {dx \over ds}(q,s) = P_{,T'} = 2 T' &;& x(q,0) = q \\ {dt \over
ds}(q,s) = P_{,\dot{T}} = -2 \dot{T} &;& t(q,0) = 0 \\ {dT \over ds}(q,s) = T' P_{,T'} + \dot{T} P_{,\dot{T}} = 2 T'^2 - 2 \dot{T}^2 &;& T(q,0) = T_i(q) \\ {dT' \over
ds}(q,s) = -P_{,x} - T' P_{,T} = 0 &;& T'(q,0) = T_{i,q}
\\ {d\dot{T} \over ds}(q,s) = -P_{,t} - \dot{T} P_{,T} = 0
&;& \dot{T}(q,0) = \pm \sqrt{1 + T_{i,q}^2 } \ .
\end{eqnarray}
There are two roots for the initial conditions equation for $\dot{T}$ because the original differential equation is expressed in terms of squares and has two branches. The asymptotic behavior of the tachyon field corresponds to $\dot{T}>0$, so from here on we will only consider that root.

The equations for $T'$ and $\dot{T}$ trivially give \begin{eqnarray} \label{tgradient} T'(q,s) &=& T_{i,q} \\ \dot{T}(q,s) &=& \sqrt{1 + T_{i,q}^2} \ . \end{eqnarray} Plugging these solutions into the equations for $x$, $t$, and $T$ gives \begin{eqnarray}
x(q,s) &=& q + 2 s T_{,q} \\ t(q,s) &=& -2 s \sqrt{1 + T_{i,q}^2} \\ T(q,s) &=& T_{i,q} - 2 s \ . \end{eqnarray}

\FIGURE[ht]{
\epsfxsize=.8\columnwidth
\epsfbox{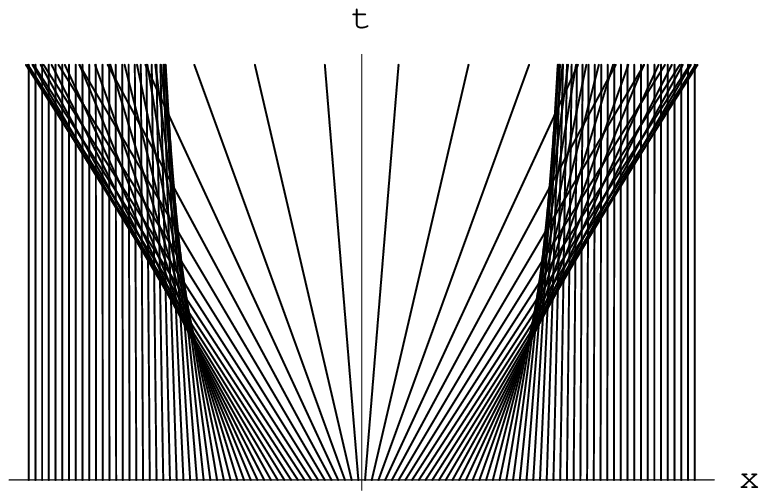}
\caption{Characteristic curves for equation (\ref{reducedsolutionx}) with $T_i = exp\left(-q^2\right)$.} \label{characteristicsfig} }

\FIGURE[ht]{
\epsfxsize=.8\columnwidth
\epsfbox{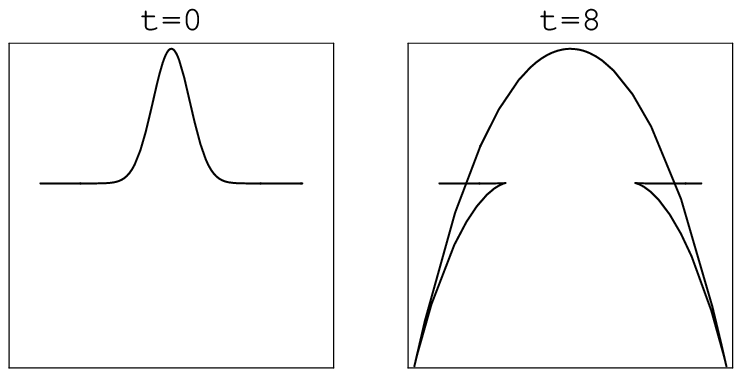}
\caption{Evolution of the field configuration $T(t, x)$ with $T_i = exp\left(-q^2\right)$.} \label{gaussevolutionfig} }

Although these equations can not be solved in the general case to eliminate both $q$ and $s$, it is a simple matter to eliminate the affine parameter $s$, which gives a parametric solution for the field $T(x(q, t), t)$ \begin{eqnarray}\label{map} \label{reducedsolutionx} x(q,t) &=& q - \frac{T_{i,q}}{\sqrt{ 1 + T_{i,q}^2}} \, t \\ \label{reducedsolutionphi} T(q,t) &=& T_i(q) + \frac{t}{ \sqrt{1 + T_{i,q}^2}} \ . \end{eqnarray} The last two equations are the main analytic results of this section.

Figures \ref{characteristicsfig} and \ref{gaussevolutionfig} illustrate how these formulas work for a particular initial profile of $T_i(q)$. Note in particular that at the moment when different characteristic curves intersect the field solution becomes multi-valued. 
Figure \ref{gauss} shows deformation and folding of the initial profile $T_i(q)$ of the realization of one dimensional 
random gaussian field.
We will have much more to say about this effect.

Note that from equation (\ref{tgradient}) we see that the gradients of the field along each characteristic always remain the same as they are at $t=0$, \begin{equation}\label{grad1}
T'(x) = T_{i,q} \ ,
\end{equation}
only spread out over a different region of space, $q \to x(t, q)$. (In the particle picture each free streaming particle maintains its initial velocity $\dot{T} = \dot{T}_i = \sqrt{1 + T_{i,q}^2}$.)  Since not too large  gradients is the condition for equation (\ref{free}) to be a good approximation to the equation of motion, this result suggests that if this approximation holds at some time $t$ it will continue to hold for all time. This argument offers an intuitive explanation for the results of our numerical simulations, namely that this asymptotic regime remains stable.

\subsection{Geometrical Interpretations}

The solution (\ref{reducedsolutionx}), (\ref{reducedsolutionphi}) represents a mapping of the initial field to the field at any time $t$ \begin{equation}\label{mapping}
T_i(q) \to T(x, t; q) \ .
\end{equation}
Here we suggest two simple geometric constructions for the hypersurface $T(x, t; q)$ at any time $t$.

\FIGURE[ht]{
\epsfxsize=.8\columnwidth
\epsfbox{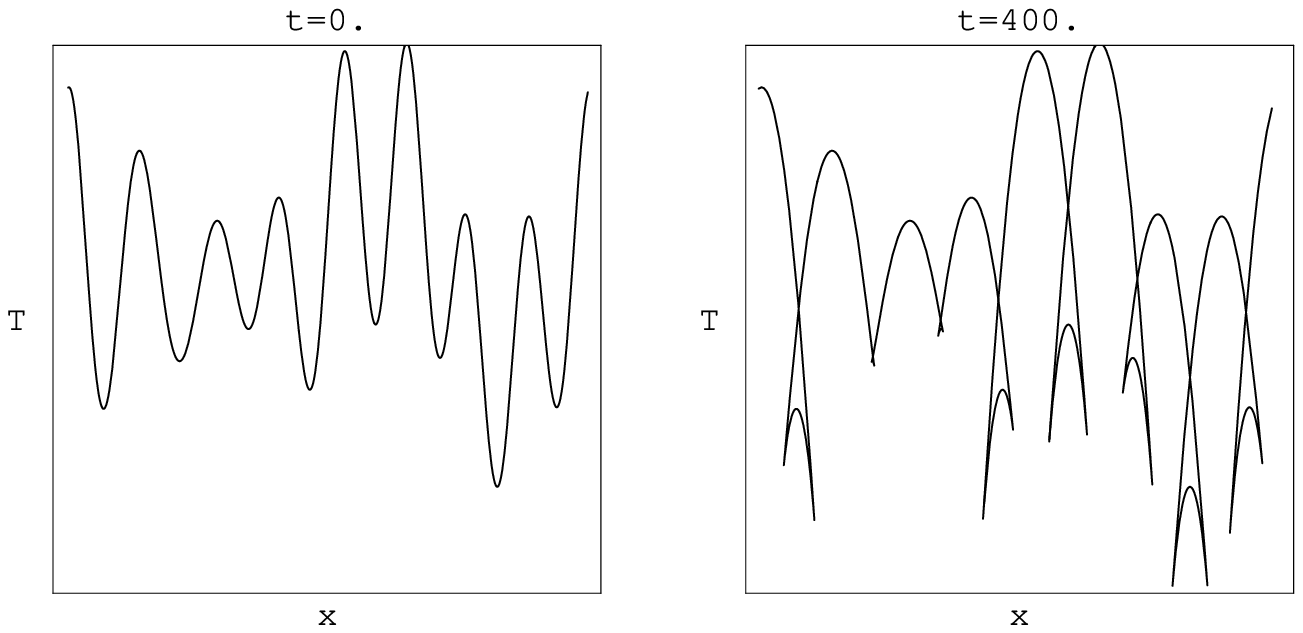}
\caption{Evolution of the field configuration $T(x, t)$ (right) with an initial realization of a random gaussian field $T_i(q)$ (left). } \label{gauss} }

The first construction allows  us to draw the deformation of the hypersurface $T$ without calculating any intermediate steps. Suppose $T_i(q)$ is that hypersurface at the initial moment $t=0$.  Consider a circle of radius equal to the time of interest $t$ and insert this circle above the curve $T_i(q)$, making it tangent to this curve at the point $q$ (without intersection), as illustrated in Fig. \ref{fig:constr1}.

\FIGURE[ht]{
\epsfxsize=.8\columnwidth
\epsfbox{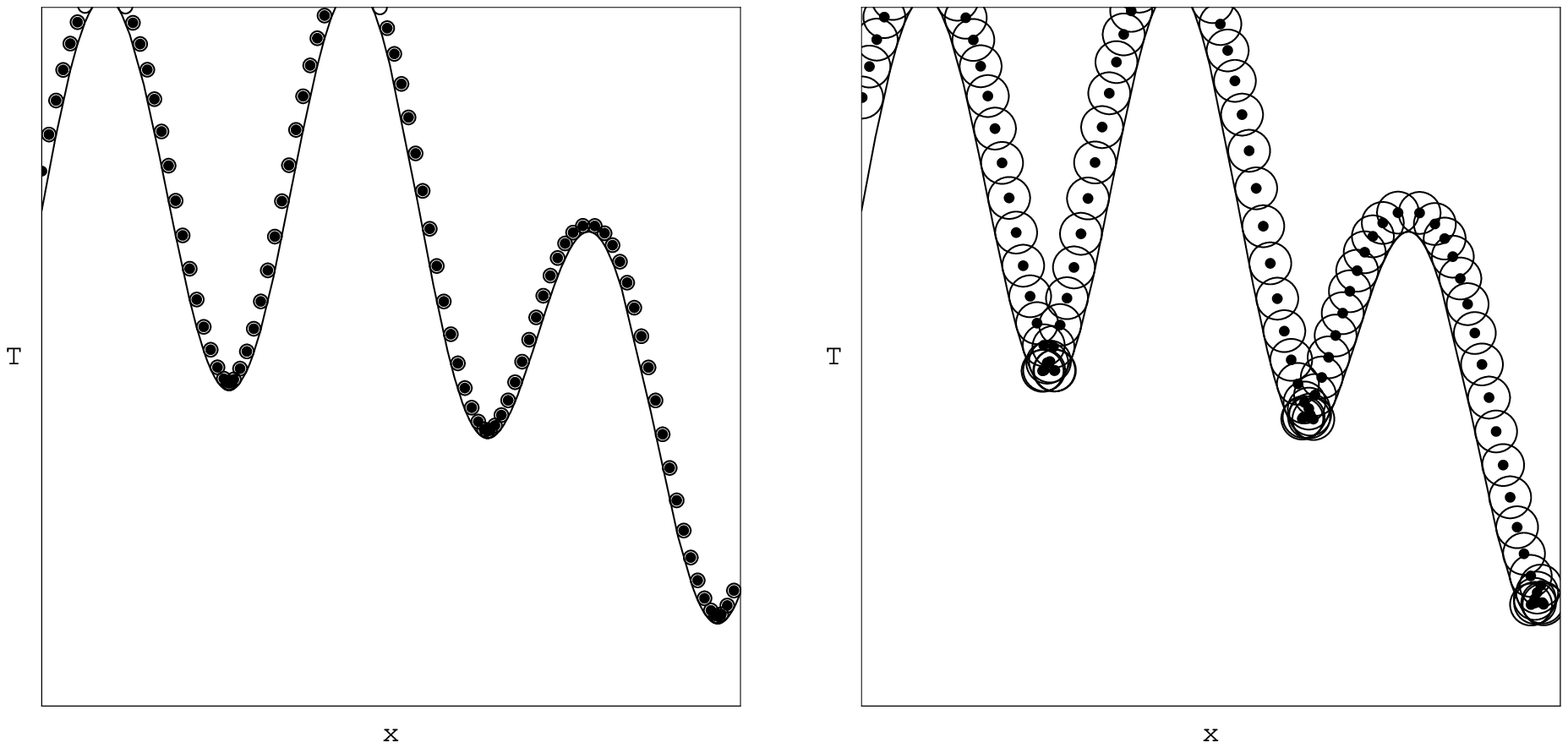}
\caption{Use of tangent circles to illustrate the deformation of the hypersurface $T(x,t;q)$.  } \label{fig:constr1} }

At the point of contact the slope of the circle is equal to the derivative  $T_{i,q}$. It is then a simple matter to calculate that the coordinates of the center of the circle will be the Eulerian coordinate $x(q,t)$ and the field value $T(x, t)$, as given by equations  (\ref{reducedsolutionx})-(\ref{reducedsolutionphi}).
 If we insert circles
of the same radius $t$ at all $q$ tangential to  $T_{i,q}$ at each point $q$, and then draw a curve through the centers of these circles, we will get the new profile of the field $T(x, t)$ at any given time. Evidently, $T(x, t)$ is flattening around the maxima of the field of $T_{i,q}$ and sharpening around its minima. At some 
moment $t_c$ there will be  a situation where a circle of radius $t_c$ is touching 
the curve $T_{i,q}$ not in one, but  in two points, $q_1$ and $q_2$. This corresponds to caustic formation. For larger $t$  this construction will decribe the folding of $T(x, t)$ with cusps, as in Figure \ref{gauss}. 

Notice that for homogeneous initial conditions $T_i(q)=const$, this construction with the circles of radius $t$ just gives the overall linear growth $T=t$.

Another, complementary geometrical construction is based on drawing the characteristics in $(t, x)$ coordinates. From each point of the initial profile $T_i(q)$ draw a straight line with slope $\frac{\sqrt{1+ T_{i,q}^2}}{T_{i,q}}$, as illustrated  in Fig. \ref{fig:constr2}. The gradient  $T'(x)$ is propagated unaltered along each of these characteristics, while the field value along each line evolves according to equation (\ref{reducedsolutionphi}).

\FIGURE[ht]{
\epsfxsize=.8\columnwidth
\epsfbox{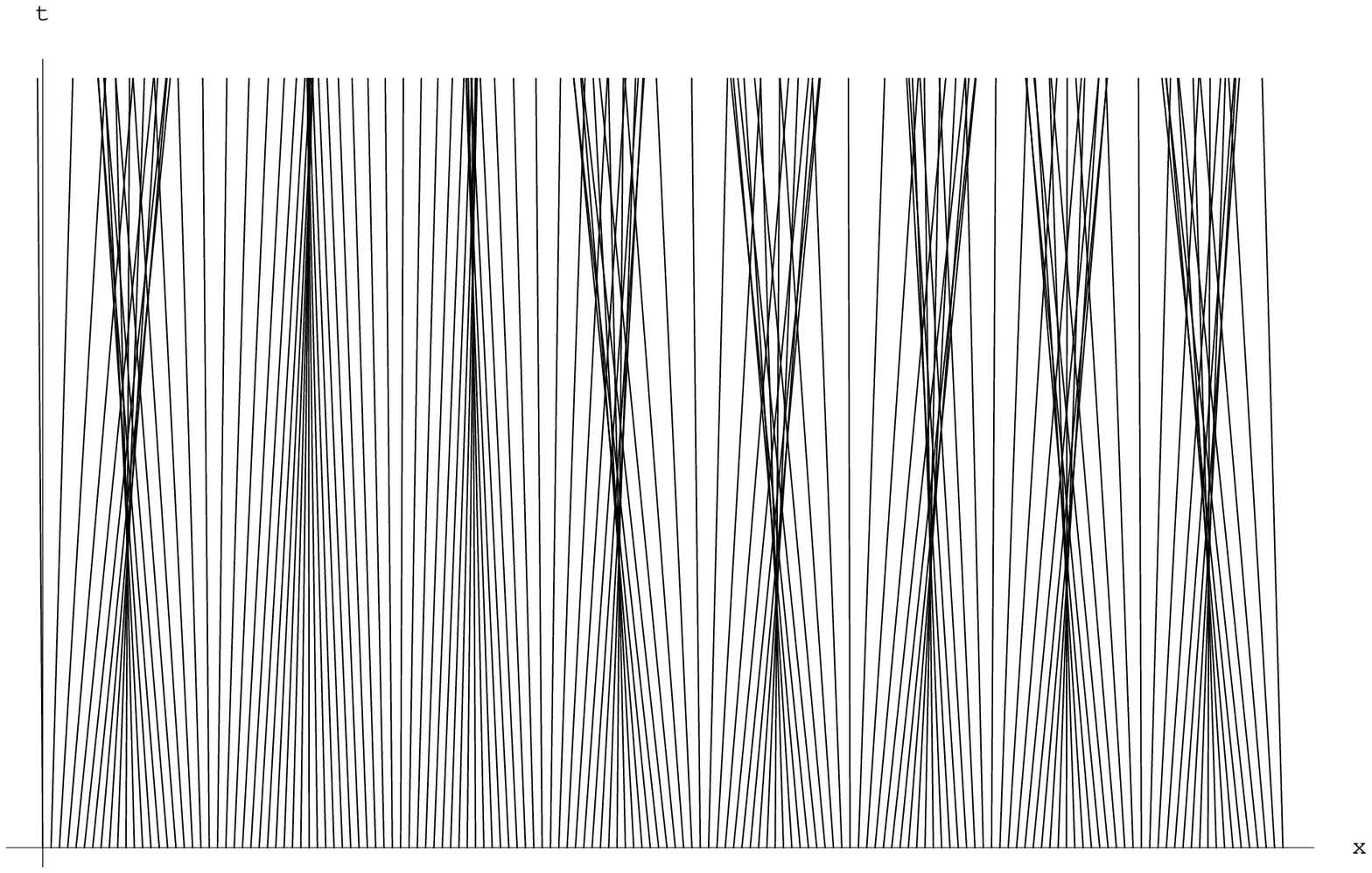}
\caption{Propagation of characteristics for the realization of initial conditions from Fig. \ref{gauss} } \label{fig:constr2} }

\subsection{Caustics and Beyond }

In this section we use the results obtained above to derive the behavior of a BI scalar with a runaway potential, i.e. one for which equation (\ref{free}) is the correct asymptotic. Consider an initial profile $T_i(q)$. The solution
(\ref{reducedsolutionx})-(\ref{reducedsolutionphi}) tells us that for each point $q$ the corresponding $x$ value will evolve linearly, as will the corresponding value of $T$. Unless $T_{i,q}$ is the same everywhere these $x$ values will move at different rates and there will eventually be different values of $q$, and thus of $T$, corresponding to the same value of $x$. In other words the field will become multi-valued. Another way of describing this behavior is to note that the characteristic curves along which our solutions are defined will in general  intersect \footnote{If $T''<0$ throughout all  space then no characteristics will ever cross, 
but this is not a physically realistic condition.}. Figure \ref{characteristicsfig} shows the characteristic curves in the $(x,t)$ plane for a toy model initial configuration $T_i(q) = exp\left(-q^2\right)$.  Figure \ref{fig:constr2} shows the propagation of characteristics for a realization of a random Gaussian field.

Since the evolution of $T$ is defined along each of these curves, the regions where they intersect are regions where this solution is multi-valued. Figure \ref{gaussevolutionfig} shows the evolution of the field configuration for the case of Figure \ref{characteristicsfig}.  Correspondingly, Figure \ref{gauss} shows it for the initial conditions of Figure \ref{fig:constr2}. In the terminology of wave front propagation, the front $T(t, x)$ deforms, and at some points it develops folding and then multi-valued regions.

The caustics are the stable singularities of the mapping (\ref{mapping}). For our case we can extend the theory of caustic formation \cite{ASZ}.  To see how they develop consider the second derivative of the field $T$, calculated from equations
(\ref{reducedsolutionx}-\ref{reducedsolutionphi})
\begin{equation}\label{second}
T''(t, x) = \left.{\partial T'(t,x) \over \partial q} \right|_t \left.{\partial q \over \partial x}\right|_t = { {T_{i,qq}} \over { 1- \frac{T_{i,qq}}{(1+T_{i,q})^{3/2}} \, t}} \ , \end{equation} where in the last step we have used the fact that $T'(x) = T_{i,q}$. At the moment of time $t_c$ and at the critical point $q_c$ where the denominator of this equations vanishes, the second derivative blows up, and a caustic forms. For a given value of $q$ this occurs at \begin{equation}\label{timec} t_c={{(1+T_{,q}(q_c))^{3/2}} \over {T_{,qq}(q_c)}}. \end{equation} Obviously, the caustics first form at the maximum of the combination $\frac{T_{,qq}}{(1+T_{,q})^{3/2}}$. From Figure \ref{fig:constr2} it is easy to see that typically the position of this maximum is well correlated with the position of the minimum of the initial field profile $T_i(q)$.

For a random gaussian field the typical time of caustic formation can be calculated in terms of the spectrum ({\it r.m.s.} values) of the gaussian field.  For simplicity, if $L$ is the typical length of small scale inhomogeneities, then we can estimate $t_c \sim L^2/T_i$.  The statistics of the caustics are defined by the statistics of the field 
$\frac{T_{i,qq}}{(1+T_{i,q})^{3/2}} \approx T_{i,qq}$.
 The number density of caustics per cubic volume will be roughly proportional to $O(10^{-2})L^{-3}$.

It is interesting to note the character of the scalar field folding for the potential $V=1$, where the action (\ref{action}) describes  a light brane in extra dimensions.  In this model the free streaming approximation does not work. However,  the field equation does admit multi-valued solutions \cite{BC} as well as caustics.  In this case a multi-valued field is simply the result of a poor choice of parametrization. For example a scalar field representing the position of a brane in a higher dimension might appear to be multi-valued if the brane folds over in an S shape. A different slicing of the extra dimension, however, will remove this problem. In the general case of runaway potentials we are considering here, however, the folding is more serious than that because the field actually crosses itself at points.

Solutions with intersecting characteristic curves -- orbit crossing -- are common in cosmology with collisionless matter. 
We will discuss this in section \ref{eqofmotion1}.

\section{The Tachyon Field in Flat  $3+1$   Dimensions}\label{dim}

Although we have been discussing the 1D case for simplicity the same basic results hold in $3+1$ dimensions, the case relevant for cosmology. In fact all of the results in this section hold for an arbitrary number of dimension. In flat spacetime the equation of motion (\ref{motion1}) reads \begin{eqnarray}\label{d3}
\ddot{T} \left(1 + ({\vec \nabla T} )^2\right)   &=& 
\left(1 -\dot{T}^2 + ({\vec \nabla T})^2\right)
\left({\vec \nabla}^2 T - {V_{,T} \over V}\right)+ \\
& +& 2 (\partial_j
\dot{T})  (\partial^j T) {\dot T} - (\partial_j \partial_k T) (\partial^j T) (\partial^k T)  \ , \end{eqnarray} where dots indicates time derivatives and $j$ and $k$ are spatial indices, $({\vec \nabla} T)^2=\partial_j T \partial^j T$. (We reserve the subscript $i$ for the initial configuration of the field $T$.)

Again, for runaway potentials the solution to equation (\ref{d3}) rapidly approaches  an asymptotic regime where it can be described by the Hamilton-Jacobi equation \begin{equation}\label{HJ}
\dot{T}^2- {\vec \nabla T}^2=1 \ ,
\end{equation}
which describes free streaming of the field $T(t, \vec x)$.

Solving this equation with the method of characteristics gives essentially the same answer that we got before, only with the Lagrangian parameter $q$ now replaced with a vector $\vec{q}$ \begin{eqnarray} \label{solution3d_1} \vec x (\vec{q},t) &=& \vec{q} - {{\nabla_{\vec q} T_i} \over {\sqrt{1 + \vert\nabla_{\vec q}T_i \vert^2}}} \, t \\
\label{solution3d_2}T(\vec{q},t) &=& T_i(\vec{q}) + {1 \over {\sqrt{1 + \vert\nabla_{\vec q}T_i \vert^2}}} \, t. \end{eqnarray}

As in the 1D case, the position and field value of each initial point move linearly with a velocity that depends on the initial gradient at that point. Thus different characteristic curves will once again intersect. However, the mapping \begin{equation}\label{mapping3} T_i(\vec q) \to T(\vec x, t; \vec q) 
\end{equation}
has a more rich structures of caustics, which form a network of two-dimensional structures \cite{ASZ}. 
The geometrical construction we used above to get the hypersurface $T(\vec x, t; \vec q)$ without step-by-step integration can be generalized to higher dimensions.  In this case we have to insert an $D$ dimensional hypersphere of radius $t$, where $D$ is the number of spatial dimensions. We place the hypersphere so that it touches the initial $D$ dimensional hypersurface $T_i(\vec q)$ tangentially at each point $\vec q$. (This process is easiest to picture for $D=2$.) The hypersurface $T(\vec x, t; \vec q)$ then will be an envelope of all centers of the inserted hyperspheres. The caustics will first form when there is a hypersphere of radius $t_c$ that simultaneously touches two points, $\vec q_1$ and $\vec q_2$.  In this case the equation (\ref{solution3d_1}) has two roots.  However, for $D>1$ there will be situations where the hypersphere touches three or more points $\vec q$. This corresponds to the merging of caustics and the formation of a network.

Let us calculate the second derivatives of the field $T$, $\nabla_j \nabla_k T$. In $3+1$ dimensions these derivatives form a three dimensional tensor.  At each point it can be diagonalized. Let $(\lambda_1, \lambda_2, \lambda_3)$ be the eigenvalues of the tensor $\nabla_j \nabla_k T$.  We also assume the ordering $\lambda_1 > \lambda_2 > \lambda_3$.  The fields $\lambda_j(\vec x, t; \vec q)$ evolve from their initial values $\lambda_{0j}(\vec q)$.  One can show that the eigenvalues propagating along the characteristics  change with time as \begin{equation}\label{lambdas} \lambda_j(\vec x, t; \vec q) = { \lambda_{0j} \over { 1- \frac{ \lambda_{0j}}{(1+T_{i,q})^{3/2}} \, t}} \ . \end{equation}

At any given point ${\vec q_c}$ a caustic will form at the time \begin{equation}\label{timec2} t_c={{(1+T_{i,q})^{3/2}} \over { \lambda_{01}(\vec q_c)}}. \end{equation} The caustics  thus first form at the maximum of the combination 
$\frac{\lambda_{01}}{(1+T_{i,q})^{3/2}}$.
For a random three dimensional random gaussian field the typical time of caustic formation can be calculated in terms of  its {\it r.m.s.} values.  The statistics of the caustics are defined by the field $\lambda_{01}(\vec q)$, which is  a non-gaussian random field.

 In some aspects our results are relativistic generalizations of the non-relativistic problem of caustic formation \cite{Zel,ASZ}.

\section{The Tachyon Field in an Expanding Universe}\label{eqofmotion1}

In this section we will discuss why consideration of the tachyon (or any other scalar) in cosmology must include consideration of inhomogeneous solutions, and we will show that for the tachyon field those solutions lead to the formation of caustics and multi-valued regions.  First we will recall some basic results for the background solution with tachyon matter and for linear gravitational instability in an expanding universe with this tachyon background \cite{FKS}. Next we will extend this analysis to the non-linear regime of gravitational instability.

\subsection{Cosmological Background of  the Rolling Tachyon}\label{backgrund}

The equation for a time-dependent homogeneous rolling tachyon in an expanding universe with a scale factor $a(t)$ follows from (\ref{motion1}) \begin{equation}\label{time} \frac{\ddot T }{1-\dot T^2}+ 3\, \frac{\dot a}{a}\, \dot T+ \frac{V_{,T}}{V} = 0\ . \end{equation}  When rolling tachyon matter dominates in an expanding universe, the equation for the evolution of the scale factor is 
\begin{equation}\label{Friedman}
{ \dot a^ 2 \over a^2 } = { 8\pi G
\over 3} { V(T) \over \sqrt { 1- \dot T^2 }} \ .
\end{equation}
 For the exponential potential of interest $V=V_0~ e^{-T/T_0}$, the homogeneous background asymptotic solution of equation (\ref{time}) 
 in an expanding universe 
for $t \gg T_0$ is \cite{FKS} 
\begin{equation}\label{f1}
T(t)= t + \frac{T_0}{4}\left({a\over a_0}\right)^6 e^{-2t/T_0} \ . \end{equation} This solution is valid for arbitrary $a(t)$, not  just when the tachyon dominates. For this solution, the energy density redshifts like ordinary matter \begin{equation}\label{energy} \varepsilon ={V(T) \over \sqrt{1- \dot T ^2 }} = V_0 \left({a_0\over a}\right)^3 \ , \end{equation} while  the pressure vanishes 
\begin{equation}\label{pressure}
p=-V(T) \sqrt {1- \dot T ^2 } = -\left({a\over a_0}\right)^3{V^2 \over V_0} \to 0 . 
\end{equation}
 When the tachyon dominates $a(t)=a_0 t^{2/3}$. 

Now we consider the case of a general potential $V(T)$.
Let us write down  $T(t)=t + \theta(t), ~|\theta|\ll t$ (note that 
$\dot \theta <0$).  To first order in $\theta$, Eq. (\ref{time}) 
reduces to
\begin{equation}
{\ddot\theta\over \dot \theta}= 2 \left.{V'\over V}\right|_{T=t}  +6{\dot a\over a} \ . \end{equation} This equation can be integrated and we obtain \begin{equation}
\dot \theta = -{1\over 2}\left({a\over a_0}\right)^6\left.{V^2\over V_0^2}\right|_{T=t}   \ ,~~~ 
1-\dot T^2 = 2\dot \theta \ ,  \label{dotth} 
\end{equation}
so the same expressions (\ref{energy}) and (\ref{pressure}) follow for $\varepsilon$ and $p$ (with the potential $V(T)$ estimated at $T=t$). 
Now, the condition $|\theta|\ll t$ requires $a^3 V\to 0$ as $t\to \infty$. If the tachyon itself dominates expansion of the Universe, this condition reduces to $T^2V(T)\to 0$ at $T\to\infty$. 
Thus, for this wide class of potentials the tachyon
 behaves  like dust-like matter at late times.

\subsection{ Linear Fluctuations in Tachyon Matter}\label{linear}

Next we consider 
small inhomogeneous perturbations of the 
tachyon field $\delta T(t, \vec{x})$ around  the time-dependent background solution 
$T(t)$ of equation (\ref{time}):
\begin{equation}\label{small}
T(t, \vec{x}) = T(t) + \delta T(t, \vec{x}) .
\end{equation}
Small scalar metric perturbations around
a FRW background can be written in the longitudinal gauge as \begin{equation}\label{adiab} ds^{2}= -\left( 1+2\Phi \right) dt^{2}+\left( 1-2\Phi \right) a^{2}\left(
t\right) d{\vec x}^2 \ .
\end{equation}
Solving the linearized Einstein equations and linearized tachyon equation of motion, we find \cite{FKS} the time evolution of the fluctuations $\Phi$ and $\delta T$ \begin{equation}\label{const} \Phi(t, \vec x)={const} \ , \hspace{1em} \delta T(t, \vec{x}) = \Phi( \vec x) \cdot t \ . \end{equation} Linear metric fluctuations are constant,  as in the cold dark matter scenario. However, the fluctuations in the tachyon field  grow.  The  tachyon fluctuations are unstable due to the effects of gravitational instability in an expanding universe.

The linear approximation for the rolling tachyon/gravity system works only during a short time interval (of the order of tens of $T_0$). Indeed, let us inspect the energy density of tachyon matter in the model with the exponential potential not assuming it to be homogeneous and using the perturbed space-time metric (\ref{adiab}): \begin{equation}\label{density1}  \varepsilon = {{V_0 e^{-T/T_0}} \over \sqrt{1-(1-2\Phi)\dot T^2 + a^{-2}  (\nabla_{\vec x}T)^2}} \ . \end{equation} For fluctuations of $\delta T$, we have $\delta T \simeq \Phi(\vec x)\, t$ where $\Phi(\vec x)$ describes the initial spatial profile of 
fluctuations. The full tachyon field including fluctuations is $T(t, \vec x)=t+(T_0/4)(a/a_0)^6 e^{-2t/T_0} + \Phi(\vec x) \,t$. The numerator of 
the expression (\ref{density1}) vanishes as $e^{-t/T_0}$, while the denominator evolves as $(a/a_0)^6 e^{-2t/T_0} + a^{-2}t^2(\nabla_{\vec x} 
\Phi)^2$. The linear approximation works during a very short time interval while $\left(\nabla_{\vec x} \Phi \right)^2 \lesssim e^{-2t/T_0}$. When this inequality breaks, linear analysis becomes insufficient. For cosmological fluctuations $\Phi \sim 10^{-5}$, the linear theory for $\delta T$ is valid during a time interval of order of $10\, T_0$.

\subsection{Non-linear Tachyon Inhomogeneities}\label{quasi}

One can try to use a quasi-linear  ansatz for inhomogeneities of $T(t,\vec x)$. 
Non-vanishing energy density (\ref{density1}) for  the rolling tachyon with 
the numerator $V\sim e^{-t/T_0}$ is achievable  only if the denominator vanishes as 
$e^{-t/T_0}$, too.  We therefore suggest the following ansatz for the inhomogeneous tachyon field \begin{equation}\label{anz} \ T\left(t, {\vec x } \right)= t + U({\vec x, t}) + f({\vec x, t}) \left({a\over a_0}\right)^6e^{-2t/T_0} \ , \end{equation} where $f({\vec x, t})$ and $U({\vec x, t})$ are two as yet undetermined functions.  In the homogeneous case $f({\vec x, t}) = const$.  We assume that in the inhomogeneous case $f({\vec x, t})$ remains sub-exponential.

Notice that
the form (\ref{anz}) corresponds to a standard procedure for obtaining the non-relativistic 
Hamilton-Jacobi equation.
 Indeed, 
the effective 4-velocity of tachyon matter is defined as \begin{equation}\label{veloc} u_{\mu }=-\frac{\nabla_{\mu} T}{(- \nabla_{\alpha}T \nabla^{\alpha}T)^{1/2}} 
\end{equation}
In the quasi-linear approximation, where  the
 3-velocity of tachyon matter is small compared to the speed of light  $c=1$, so that $u^0=-u_0\to 1$ for $t\to\infty$, we have 
\begin{equation}\label{veloc1}
u_{\mu } \approx  - \nabla_{\mu} T \ .
\end{equation}

 The most interesting part of the expression (\ref{anz}) is the term containing the function 
  $U({\vec x, t})$, which survives in late time asymptotics. 
 To see the effect of this term, we substitute the form (\ref{anz}) in the Einstein equations. We further assume that the gravitational potential is still linear (as in 
 the usual theory of gravitational
instability with ordinary matter), while inhomogeneities in tachyon field are not. Then from the Einstein equations, where we keep  the gravitational potential linearized, we derive \begin{equation}\label{Bern} \dot U - {1\over 2a^2}(\nabla_{\vec x}U)^2  =\Phi  \ . \end{equation} This equation  is supplemented by the Poisson equation for the gravitational potential, \begin{equation}\label{Pois} \nabla_{\vec x}^2\Phi = 4\pi G \, a^2 (\varepsilon (t,\vec x)-\bar\varepsilon (t)) \ , \end{equation}  which, however,  we will not use here. In the linear regime this equation gives us back the right linear expression 
$\delta T= U=\Phi  t$.
Eq. (\ref{Bern}) is similar to the cosmic Bernoulli equation for dust-like matter with $U$ being a (minus) scalar velocity potential for the 
  effective 3-velocity of tachyon matter.

It is easy to recognize that Eq. (\ref{Bern})
comes from the covariant form of the Hamilton-Jacobi equation \begin{equation}\label{CHJ} g^{\mu\nu} \partial_{\mu} S \, \partial_{\nu} S=1 \ . \end{equation} Indeed, for the metric (\ref{adiab}) we have $g^{00}=-(1-2\Phi)$, $g^{ij}={1 \over a^2}(1+2\Phi)\delta^{ij}$. Substituting these in (\ref{CHJ}), we get \begin{equation}\label{inter} (1-2\Phi)\left(\frac{\partial S}{  \partial t}\right)^2-{1 \over a^2}(1+2\Phi) \left(\nabla_{\vec x} S \right)^2= 1 \ . \end{equation} For  a linear gravitational potential $\Phi$ we  have \begin{equation}\label{CHJ1} \dot S^2 -{1\over a^2}(\nabla_{\vec x}S)^2=1+2\Phi \ . \end{equation} The form (\ref{anz}) for $S$, $S \approx t+U$,  immediately leads to equation 
(\ref{Bern}) for $U$.

 This suggest that we can go even further beyond the quasi-linear ansatz (\ref{anz}) to a fully non-linear, covariant 
free streaming approximation (\ref{CHJ}).
 Equation
(\ref{CHJ1}) describes free wave front propagation in a gravitational  field described by the gravitational potential $\Phi$. Equation (\ref{CHJ})  describes this propagation in a general curved space time. The common feature of the wave front propagation in all these settings is the formation of caustics,  which occurs similarly to how we described it in the previous sections.

In application to tachyon cosmology, we can continue to use the geometrical  techniques we developed for  flat spacetime. However, in  the cosmological problem the initial hypersurface of $T(\vec q)$ is defined by the linear stage of gravitational instability 
and is given by the initial profile of the scalar metric perturbations, $T(\vec q)= \Phi(\vec q) t_0$, where $t_0$ is some initial time moment. Therefore the timing of appearance and statistics of caustics  are defined by  the scales, amplitude and statistics of the initial metric perturbations $\Phi(\vec q)$,  as in the problem of 
gravitational clustering in the standard cold dark matter scenario. We expect caustics in the tachyon matter to form around the moment of equilibrium $z_{eq}$ between radiation and matter, when 
inhomogeneities start growing in the matter dominated regime. The density of caustics will be defined by the minimal scale of the cosmological fluctuations, say, $1$ mm (in present day  scales).

Caustics in the theory of cosmological gravitational instability are common for  models with cold collisionless matter.  In the theory of the large scale structure of the universe they are described by the famous Zel'dovich approximation \cite{Zel}.  The theory of caustics in pressureless media has been well developed \cite{ASZ} and includes many dynamical and statistical aspects \cite{SZ} and geometrical techniques for constructing the mapping $q \to x(q, t)$ \cite{GSS}. The principal difference, however, is that for media of particles the physics is well defined beyond the caustics, while here we are dealing with a non-linear scalar field whose action is ill-defined in the multi-valued regime.  Another important difference is that caustics in media correspond to infinite values of mass density, while it is not clear whether 
  the energy density
$\varepsilon$ will be singular at caustics  
in the case of a BI scalar field. 

\section{Conclusion}\label{discussion}

We considered field theory with the action (\ref{action}).
At first glance  the equation of motion for  the evolution of  the non-linear inhomogeneous field $T(t, \vec x)$ looks very complicated. Remarkably, we found that in  flat spacetime for 
several examples of  runaway potentials $V(T)$
with the ground state at infinity, generic solutions 
of the theory (\ref{action}) very quickly approach an asymptotic form $T(t, \vec x) \to S(t, \vec x) $, which obeys the Hamilton-Jacobi equation \begin{equation} \dot S^2 -({\vec \nabla_{\vec x}}S )^2=1 \ . \end{equation} Considering an example of  the theory 
(\ref{action}) in an expanding universe with small metric perturbations,  we found that in this more complicated case  the equation of the 
tachyon  is again very well approximated by the solution of the covariantly generalized Hamilton-Jacobi equation \begin{equation}\label{Ham} g^{\mu\nu} \frac{\partial S}{ \partial x^{\mu}} \frac{\partial S}{ \partial x^{\nu}} =1 \ . \end{equation} We conjecture that the solution $S$ of the Hamilton-Jacobi equation 
(\ref{Ham}) is an attractor for the equation of motions of the theory 
(\ref{action}) with a runaway potentail $V(T)$.
It will be interesting to check if this conjecture can be extended to  other non-trivial geometries $g_{\mu\nu}$ and how it depends on $V(T)$. 

Another question we  would like to address is the character of convergency of the solution of the actual tachyon equation to the approximated function $S$ which obeys (\ref{Ham}).  Generalizing the convergence of the homogeneous 
runaway tachyon solution, we may suggest the asymptotic expansion \begin{equation}\label{asymp} T = S + f_1 \, e^{-S}+f_2 \,  e^{-2S} + ... \ , \end{equation} where $f_1, f_2, ...$ are non-exponential functions of $t$ and $x$. 
The functions $f_1, f_2, ...$ can  be important, when we address the issue about the value of energy density at caustics.

 The most important finding of our study is
that generic solutions of theory (\ref{action}) with a runaway potential contain  caustics and multi-valued regions of $T$. 
The interpretation of the multi-valued field $T$ may depend on the potential $V(T)$ and the dimensionality $D$ of the theory. However, the meaning of the  multi-valued solutions in an effective four dimensional theory with a runaway potential is not clear to us. 

The model (\ref{action}) can be used as an approximation
 to the string theory tachyon when higher order derivatives can be neglected \begin{equation}\label{aaa} S=-\int d^4x \sqrt{-g} V(T) \sqrt{1+\alpha' \nabla_{\mu} T \nabla^{\mu}T} + O(\nabla_{\mu}\nabla^{\mu} T) \ . \end{equation} We  found, however,  that the generic solution of the equations where we keep only  low order gradients is not consistent  because at the caustics second and higher derivatives blow up. Therefore the simple theory (\ref{action}) does not work for the string theory tachyon. As a result, at this point,  we can say little about cosmological  applications of the tachyon.

The situation is different for the theory (\ref{aaa}) 
with the potential $V(T)$ which has a ground state at finite $T$. 
In this case caustics may not develop. 
However, in this case, as was shown in \cite{FKS}, the tachyon field model (\ref{action}) describes neither non-relativistic cold dark matter nor dark energy in the 
present Universe.

Another aspect that should be included in  a comprehensive  analysis of the tachyon is its interaction with  the dilaton field, gauge fields and  others \cite{sigma}. It will be interesting to see if the interaction is especially important on caustics 
or if the backreaction of such  interactions may even affect their formation.

\section*{Acknowledgements}
We would like to thank Andrei Frolov,  Renata Kallosh,
 Andrei Linde and Sergei Shandarin for discussions and comments. We thank 
Richard Easther for 
 discussions and for  help with our numerical
simulations. We thank
NATO Linkage Grant 975389 for support. G.F. was supported by PREA, L.K. was supported by NSERC and CIAR. The work of A.S. in Russia was partially supported by RFBR, grants No 02-02-16817 and 00-15-96699, and by the RAS Research Program ``Quantum Macrophysics''. A.S. thanks CITA for hospitality during his visit.

\end{document}